\documentclass{article}
\usepackage[latin1]{inputenc}%
\usepackage[pdftex]{graphicx}
\usepackage[colorlinks=false, pdfborder={0 0 0}]{hyperref}%
\DeclareGraphicsExtensions{.pdf}%

\usepackage{geometry}
\usepackage[sort&compress]{natbib}
\newcommand{\image}{\mathbf{I}} 
\newcommand{\dico}{\Phi} 

\newcommand{\coef}{\mathbf{a}} 

\newcommand\la{\leftarrow} %
\newcommand{\sparsenet}{{\sc SparseNet}}%
\newcommand{\seeFig}[1]{figure~\ref{fig:#1}}%
\newcommand{\seeSec}[1]{section~\ref{sec:#1}}%
\newcommand{\seeEq}[1]{equation~\ref{eq:#1}}%
\usepackage{units}%
\usepackage{fancyhdr}
\date{July 2010}
\hypersetup{%
 pdftitle={%
Role of homeostasis in learning sparse representations},%
 pdfsubject={ $Id: perrinet10shl.tex 4505 2011-01-27 07:25:43Z perrinet $ },%
 pdfauthor={Laurent Perrinet <Laurent.Perrinet@univ-amu.fr>, INCM/CNRS, 31, ch. Joseph Aiguier, 13402 Marseille Cedex 20, France - http://invibe.net/LaurentPerrinet},%
 pdfkeywords={Neural population coding, Unsupervised learning, Statistics of natural images, Simple cell receptive fields, Sparse Hebbian Learning, adaptive Matching Pursuit, Cooperative Homeostasis, Competition-Optimized Matching Pursuit},%
}%
\title{%
Role of homeostasis in learning sparse representations%
}%
\author{Laurent U.~Perrinet\\%
Institut de Neurosciences de la Timone (UMR7289)\\ CNRS / Aix-Marseille Universit\'e --- France\\
e-mail: \url{Laurent.Perrinet@univ-amu.fr}\\%
\url{http://invibe.net/LaurentPerrinet/Publications/Perrinet10shl} %
}%
\begin{document}%
\maketitle %
\paragraph{BibTex entry}~~\\
\begin{verbatim}
@article{Perrinet10shl,
	Author = {Perrinet, Laurent U.},
	Title = {Role of homeostasis in learning sparse representations},
	Year = {2010},
	Arxivid = {0706.3177},
	Doi = {10.1162/neco.2010.05-08-795},
	ISSN = {1530-888X},
	Journal = {Neural Computation},
	Keywords = {Neural population coding, Unsupervised learning,
		Statistics of natural images, Simple cell receptive fields, 
		Sparse Hebbian Learning, Adaptive Matching Pursuit, 
		Cooperative Homeostasis, Competition-Optimized Matching Pursuit},
	Month = {July},
	Volume = {22},
	Number = {7},
	Pages = {1812--36},
	Url = {http://invibe.net/LaurentPerrinet/Publications/Perrinet10shl},
}\end{verbatim}
\begin{abstract}%
Neurons in the input layer of primary visual cortex in primates develop edge-like receptive fields. One approach to understanding the emergence of this response is to state that neural activity has to efficiently represent sensory data with respect to the statistics of natural scenes. Furthermore, it is believed that such an efficient coding is achieved using a competition across neurons so as to generate a sparse representation, that is, where a relatively small number of neurons are simultaneously active. Indeed, different models of sparse coding coupled with Hebbian learning and homeostasis have been proposed that successfully match the observed emergent response. However, the specific role of homeostasis in learning such sparse representations is still largely unknown. By quantitatively assessing the efficiency of the neural representation during learning, we derive a cooperative homeostasis mechanism which optimally tunes the competition between neurons within the sparse coding algorithm. We apply this homeostasis while learning small patches taken from natural images and compare its efficiency with state-of-the-art algorithms. Results show that while different sparse coding algorithms give similar coding results, the homeostasis provides an optimal balance for the representation of natural images within the population of neurons. Competition in sparse coding is optimized when it is fair: By contributing to optimize statistical competition across neurons, homeostasis is crucial in providing a more efficient solution to the emergence of independent components.%
\end{abstract}%
\subsection*{Keywords}
Neural population coding, Unsupervised learning, Statistics of natural images, Simple cell receptive fields, Sparse Hebbian Learning, Adaptive Matching Pursuit, Cooperative Homeostasis, Competition-Optimized Matching Pursuit%

\section{Introduction}%
\label{sec:intro}%
The central nervous system is a dynamical, adaptive organ which constantly evolves to provide optimal decisions for interacting with the environment. The early visual pathways provide a powerful system for probing and modeling these mechanisms. For instance, it is observed that edge-like receptive fields emerge in simple cell neurons from the input layer of the primary visual cortex of primates~\citep{Chapman92}. The development of cortical cell orientation tuning is an activity-dependent process but it is still largely unknown how neural computations implement this type of unsupervised learning mechanisms. A popular view is that such a population of neurons operates so that relevant sensory information from the retino-thalamic pathway is transformed (or ``coded'') efficiently. Such efficient representation will allow decisions to be taken optimally in higher-level layers or areas~\citep{Atick92,Barlow01}. It is believed that this is achieved through lateral interactions which remove redundancies in the neural representation, that is, when the representation is \emph{sparse}~\citep{Olshausen96a}. A representation is sparse when each input signal is associated with a relatively small number of simultaneously activated neurons within the population. For instance, orientation selectivity of simple cells is sharper that the selectivity that would be predicted by linear filtering. As a consequence, representation in the orientation domain  is sparse and allows higher processing stages to better segregate edges in the image~\citep{Field94}. Sparse representations are observed prominently with cortical response to natural stimuli, that is, to behaviorally relevant sensory inputs~\citep{Vinje00,Deweese03,Baudot04}. This reflects the fact that, at the learning time scale, coding is optimized relative to the statistics of natural scenes.  The emergence of edge-like simple cell receptive fields in the input layer of the primary visual cortex of primates may thus be considered as a coupled coding and learning optimization problem: At the coding time scale, the sparseness of the representation is optimized for any given input while at the learning time scale, synaptic weights are tuned to achieve on average optimal representation efficiency over natural scenes.%

Most of existing models of unsupervised learning aim at optimizing a cost defined on prior assumptions on representation's sparseness. These sparse learning algorithms have been applied both for images~\citep{Fyfe95,Olshausen96a,Zibulevsky01,Perrinet04,Rehn07,Doi07} and sounds~\citep{Lewicki00,Smith06}. For instance, learning is accomplished in \sparsenet~\citep{Olshausen96a} on patches taken from natural images as a sequence of coding and learning steps. First, sparse coding is achieved using a gradient descent over a convex cost derived from a sparse prior probability distribution function of the representation. At this step of the learning, it is performed using the current state of the ``dictionary'' of receptive fields. Then, knowing this sparse solution, learning is defined as slowly changing the dictionary using Hebbian learning. In general, the parameterization of the prior has major impacts on results of the sparse coding and thus on the emergence of edge-like receptive fields and requires proper tuning. In fact, the definition of the prior corresponds to an objective sparseness and does not always fit to the observed probability distribution function of the coefficients. In particular, this could be a problem \emph{during} learning if we use the cost to measure representation efficiency for this learning step. An alternative is to use a more generic L$_0$ norm sparseness, by simply counting the number of non-zero coefficients. It was found that by using an algorithm like Matching Pursuit, the learning algorithm could provide results similar to \sparsenet , but without the need of parametric assumptions on the prior~\citep{Perrinet03,Perrinet04,Smith06,Rehn07}. However, we observed that this class of algorithms could lead to solutions corresponding to a local minimum of the objective function: Some solutions seem as efficient as others for representing the signal but do not represent edge-like features homogeneously. In particular, during the early learning phase, some cells may learn ``faster'' than others. There is the need for a homeostasis mechanism that will ensure convergence of learning. The goal of this work is to study the specific role of homeostasis in learning sparse representations and to propose a homeostasis mechanism which optimizes the learning of an efficient neural representation.%

To achieve this, we first formulate analytically the problem of representation efficiency in a population of sensory neurons (\seeSec{method})  and define the class of Sparse Hebbian Learning (SHL) algorithms. For the particular non-parametric L$_0$ norm sparseness, we show that sparseness is optimal when average activity within the neural population is uniformly balanced. Based on a previous implementation, Adaptive Matching Pursuit (AMP)~\citep{Perrinet03,Perrinet04}, we will define a homeostatic gain control mechanism that we will integrate in a novel SHL algorithm~(\seeSec{ssc}). Finally, we compare in \seeSec{results} this novel algorithm with AMP and the state-of-the-art \sparsenet\ method~\citep{Olshausen96a}. Using quantitative measures of efficiency based on constraints on the neural representation, we show the importance of the homeostasis mechanism in terms of representation efficiency. We conclude in \seeSec{discussion} by linking this original method with other Sparse Hebbian Learning schemes and how these may be united to improve our understanding of the emergence of edge-like simple cell receptive fields, drawing the bridge between structure (representation in a distributed network) and function (efficient coding).%
\section{Problem Statement}%
\label{sec:method}%
\subsection{Definition of representation efficiency}%
\label{sec:lgm}%
In low-level sensory areas, the goal of neural computations is to generate efficient intermediate \emph{representations} to allow efficient decision making. Classically, a representation is defined as the inversion of an internal generative model of the sensory world, that is, by inferring the sources that generated the input signal. Formally, as in~\citet{Olshausen98}, we define a Linear Generative Model (LGM) for describing natural, static, grayscale images $\image$ (represented by column vectors of dimension $L$ pixels), by setting a ``dictionary'' of $M$ images (or ``filters'') as the $L \times M$ matrix $\dico=\{ \dico_i\}_{1\leq i \leq M}$. Knowing the associated ``sources'' as a vector of coefficients $\coef=\{ a_i \}_{1\leq i \leq M}$, the image is defined using matrix notation as%
\begin{equation}%
\image = \dico\coef + \mathbf{n}%
\label{eq:lgm}%
\end{equation} %
where $\mathbf{n}$ is a decorrelated gaussian additive noise image of variance $\sigma_n^2$. The decorrelation of the noise is achieved by applying Principal Component Analysis to the raw input images, without loss of generality since this preprocessing is invertible. Generally, the dictionary $\dico$ may be much larger than the dimension of the input space (that is, if $M \gg L$) and it is then said to be \emph{over-complete}. However, given an over-complete dictionary, the inversion of the LGM leads to a combinatorial search and typically, there may exist many coding solutions $\coef$ to \seeEq{lgm} for one given input $\image$. The goal of efficient coding is to find, given the dictionary $\dico$ and for any observed signal $\image$, the ``best'' representation vector, that is, as close as possible to the sources that generated the signal. It is therefore necessary to define an efficiency criterion in order to choose between these different solutions.%

Using the LGM, we will infer the ``best'' coding vector as the most probable. In particular, from the physical synthesis of natural images, we know \emph{a priori} that image representations are sparse: They are most likely generated by a small number of features relatively to the dimension $M$ of representation space. Similarly to \citet{Lewicki00}, this can be formalized in the probabilistic framework defined by the LGM (see~\seeEq{lgm}), by assuming that we know the prior distribution of the coefficients $a_i$ for natural images. The representation cost of $\coef$ for one given natural image is then:%
\begin{eqnarray}%
\mathcal{C}( \coef | \image , \dico) 	&=& -\log P( \coef | \image , \dico) \nonumber \\
&=& \log Z + \frac{1}{2\sigma_n^2} \| \image - \dico \coef \|^2 - \sum_i \log P(a_i | \dico)%
\label{eq:efficiency_cost}%
\end{eqnarray}%
where $Z$ is the partition function which is independent of the coding and $ \| \cdot \|$ is the L$_2$ norm in image space. This efficiency cost is measured in bits if the logarithm is of base 2, as we will assume without loss of generality thereafter. For any representation $\coef$, the cost value corresponds to the description length~\citep{Rissanen78}: On the right hand side of~\seeEq{efficiency_cost}, the second term corresponds to the information from the image which is not coded by the representation (reconstruction cost) and thus to the information that can be at best encoded using entropic coding pixel by pixel (it's the log-likelihood in Bayesian terminology). The third term $S( \coef | \dico) = - \sum_i \log P(a_i | \dico)$ is the representation or sparseness cost: It quantifies representation efficiency as the coding length of each coefficient of $\coef$ independently which would be achieved by entropic coding knowing the prior. 
In practice, the sparseness of coefficients for natural images is often defined by an \emph{ad hoc} parameterization of the prior's shape. For instance, the parameterization in~\citet{Olshausen98} yields the coding cost:%
\begin{equation}
\mathcal{C}_1( \coef | \image , \dico) =\frac{1}{2\sigma_n^2} \| \image - \dico \coef \|^2 + \beta \sum_i \log ( 1 + \frac{a_i^2}{\sigma^2} )%
\label{eq:sparse_cost}%
\end{equation}
where $\beta$ corresponds to the prior's steepness and $\sigma$ to its scaling (see Figure 13.2 from~\citep{Olshausen02}). This choice is often favored because it results in a convex cost for which known numerical optimization methods such as conjugate gradient may be used. 

A non-parametric form of sparseness cost may be defined by considering that neurons representing the vector $\coef$ are either active or inactive. In fact, the spiking nature of neural information demonstrates that the transition from an inactive to an active state is far more significant at the coding time scale than smooth changes of the firing rate. This is for instance perfectly illustrated by the binary nature of the neural code in the auditory cortex of rats~\citep{Deweese03}. Binary codes also emerge as optimal neural codes for rapid signal transmission~\citep{Bethge03,Nikitin09}. With a binary event-based code, the cost is only incremented when a new neuron gets active, regardless to the analog value. Stating that an active neuron carries a bounded amount of information of $\lambda$ bits, an upper bound for the representation cost of neural activity on the receiver end is proportional to the count of active neurons, that is, to the L$_0$ norm:%
\begin{equation}%
\mathcal{C}_0( \coef | \image , \dico) = \frac{1}{2\sigma_n^2} \| \image - \dico \coef \|^2 + \lambda \| \coef \|_0%
\label{eq:L0_cost}%
\end{equation}%
This cost is similar with information criteria such as the AIC~\citep{Akaike74} or distortion rate~\citep[p.~571]{Mallat98}. This simple non-parametric cost has the advantage of being dynamic: The number of active cells for one given signal grows in time with the number of spikes reaching the receiver (see architecture of the model in~\seeFig{laughlin}-Left). But \seeEq{L0_cost} defines a harder cost to optimize since the hard L$_0$ norm sparseness leads to a non-convex optimization problem which is \emph{NP-complete} with respect to the dimension $M$ of the dictionary~\citep[p.~418]{Mallat98}.%
\subsection{Sparse Hebbian Learning (SHL)}%
\label{sec:shl}%
Given a sparse coding strategy that optimizes any representation efficiency cost as defined above, we may derive an unsupervised learning model by optimizing the dictionary $\dico$ over natural scenes. On the one hand, the flexibility in the definition of the sparseness cost leads to a wide variety of proposed \emph{sparse coding} solutions (for a review, see~\citep{Pece02}) such as numerical optimization~\citep{Olshausen98,Lee07}, non-negative matrix factorization~\citep{Lee99,Ranzato07} or Matching Pursuit~\citep{Perrinet03,Perrinet04,Smith06,Rehn07}. On the other hand, these methods share the same LGM model (see~\seeEq{lgm}) and once the sparse coding algorithm is chosen, the learning scheme is similar.%

Indeed, after every coding sweep, the efficiency of the dictionary $\dico$ may be increased with respect to \seeEq{efficiency_cost}. By using the online gradient descent approach given the current sparse solution, learning may be achieved using $\forall i$:
\begin{equation}%
\dico_{i} \la \dico_{i} + \eta a_{i} (\image - \dico\coef)%
\label{eq:learn}%
\end{equation}%
where $\eta$ is the learning rate. %
Similarly to Eq.~17 in~\citep{Olshausen98} or to Eq.~2 in~\citep{Smith06}, the relation is a linear ``Hebbian'' rule~\citep{Hebb49} since it enhances the weight of neurons proportionally to the correlation between pre- and post-synaptic neurons. Note that there is no learning for non-activated coefficients. The novelty of this formulation compared to other linear Hebbian learning rule such as~\citep{Oja82} is to take advantage of the sparse representation, hence the name Sparse Hebbian Learning (SHL).%

SHL algorithms are unstable without homeostasis. In fact, starting with a random dictionary, the first filters to learn are more likely to correspond to salient features~\citep{Perrinet03ieee} and are therefore more likely to be selected again in subsequent learning steps. In {\sc SparseNet}, the homeostatic gain control is implemented by adaptively tuning the norm of the filters. This method equalizes the variance of coefficients across neurons using a geometric stochastic learning rule. The underlying heuristic is that this introduces a bias in the choice of the active coefficients. In fact, if a neuron is not selected often, the geometric homeostasis will decrease the norm of the corresponding filter, and therefore ---from~\seeEq{lgm} and the conjugate gradient optimization--- this will increase the value of the associated scalar. Finally, since the prior functions defined in~\seeEq{sparse_cost} are identical for all neurons, this will increase the relative probability that the neuron is selected with a higher relative value. The parameters of this homeostatic rule have a great importance for the convergence of the global algorithm. We will now try to define a more general homeostasis mechanism derived from the optimization of representation efficiency.
\subsection{Efficient cooperative homeostatis in SHL}%
\label{sec:coding} %
\begin{figure}
\includegraphics[width=\textwidth]{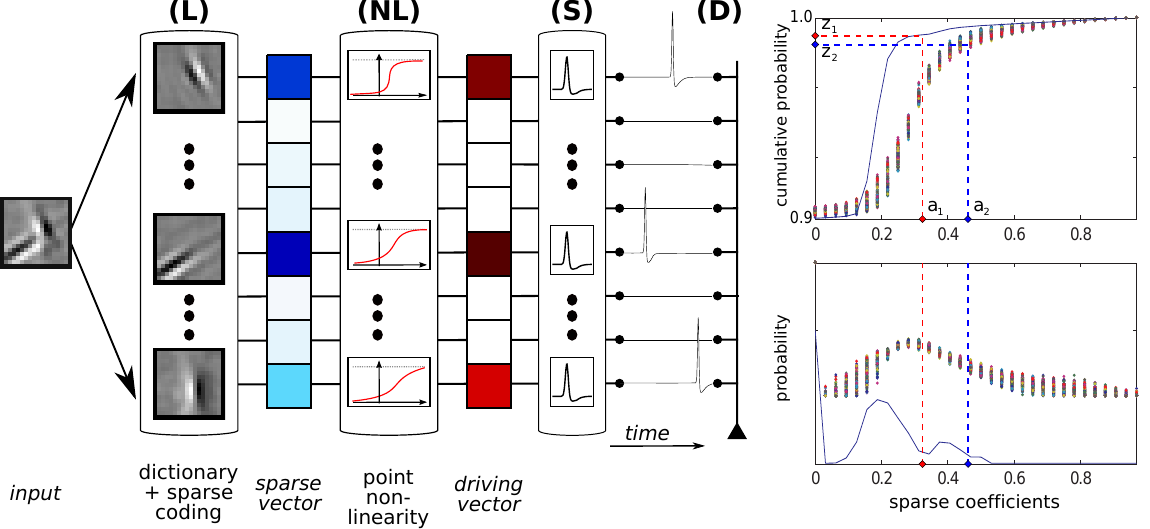}%
\caption{\textbf{Simple neural model of sparse coding and role of homeostasis.} \textit{(Left)} We define the coding model as an information channel constituted by a bundle of Linear/Non-Linear spiking neurons. \textbf{(L)} A given input image patch is coded linearly by using the dictionary of filters $\dico_i$ and transformed by sparse coding (such as Matching Pursuit) into a sparse vector $\coef$. Each coefficient is transformed into a driving coefficient in the \textbf{(NL)} layer by using a point non-linearity which \textbf{(S)} drives a generic spiking mechanism. \textbf{(D)} On the receiver end (for instance in an efferent neuron), one may then estimate the input from the neural representation pattern. This decoding is progressive, and if we assume that each spike carries a bounded amount of information, representation cost in this model increases proportionally with the number of activated neurons. \textit{(Right)} However, for a given dictionary, the distribution of sparse coefficients $a_i$ and hence the probability of a neuron's activation is in general not uniform. We show (Lower panel) the log-probability distribution function and (Upper panel) the cumulative distribution of sparse coefficients for a dictionary of edge-like filters with similar selectivity (dotted scatter) except for one filter which was randomized (continuous line). This illustrates a typical situation which may occur during learning when some components did learn less than others: Since their activity will be lower, they are less likely to be activated in the spiking mechanism and from the Hebbian rule, they are less likely to learn. When selecting an optimal sparse set for a given input, instead of comparing sparse coefficients with respect to a threshold (vertical dashed lines), it should instead be done on the significance value $z_i$ (horizontal dashed lines): In this particular case, the less selective neuron ($a_1 < a_2$) is selected by the homeostatic cooperation ($z_1 > z_2$). The role of homeostasis during learning is that, even if the dictionary of filters is not homogeneous, the point non-linearity in (NL) modifies sparse coding in (L) such that the probability of a neuron's activation is uniform across the population.%
} %
\label{fig:laughlin}%
\end{figure}
The role of homeostasis during learning is to make sure that the distribution of neural activity is homogeneous. In fact, neurons belonging to a same neural assembly~\citep{Hebb49} form a competitive network and should \emph{a priori} carry similar information. This optimizes the coding efficiency of neural activity in terms of compression~\citep{Hateren93} and thus minimizes intrinsic noise~\citep{Srinivasan82}. Such a strategy is similar to introducing an intrinsic adaptation rule such that prior firing probability of all neurons have a similar Laplacian probability distribution~\citep{Weber08}. Dually, since neural activity in the assembly actually represents the sparse coefficients, we may understand the role of homeostasis as maximizing the average representation cost $\mathcal{C}( \coef | \dico)$ at the time scale of learning. This is equivalent to say that homeostasis should act such that at any time, invariantly to the selectivity of features in the dictionary, the probability of selecting one feature is uniform across the dictionary.%

This optimal uniformity may be achieved in all generality for any given dictionary by using point non-linearities $z_i$ applied to the sparse coefficients: In fact, a standard method to achieve uniformity is to use an equalization of the histogram~\citep{Atick92}. This method may be easily derived if we know the probability distribution function $dP_{i}$ of variable $a_i$ by choosing the non-linearity as the cumulative distribution function transforming any observed variable $\bar{a}_i$ into:
\begin{equation}
z_i(\bar{a}_i)= P_{i}(a_i \leq \bar{a}_i) = \int_{-\infty}^{\bar{a}_i} dP_{i}(a_i)%
\label{eq:laughlin}%
\end{equation}
This is equivalent to the change of variables which transforms the sparse vector $\coef$ to a variable with uniform probability distribution function in $[0, 1]^M$. The transformed coefficients may thus be used as a normalized drive to the spiking mechanism of the individual neurons (see~\seeFig{laughlin}-Left). This equalization process has been observed in the neural activity of a variety of species and is, for instance, perfectly illustrated in the salamander's retina~\citep{Laughlin81}. It may evolve dynamically to slowly adapt to varying changes in luminance or contrast values, such as when the light diminishes at twilight~\citep{Hosoya05}.%

This novel and simple non-parametric homeostatic method is applicable to Sparse Hebbian Learning algorithms by using this transform on the sparse coefficients. Let's imagine for instance that one filter corresponds to a feature of low selectivity while others correspond to similarly selective features: As a consequence, this filter will correspond on average to lower sparse coefficients (see~\seeFig{laughlin}-Right). However, the respective gain control function $z_i$ will be such that all transformed coefficients have the same probability density function. Using the transformed coefficients to evaluate which neuron should be active, the homeostasis will therefore optimize the information in the representation cost defined in~\seeEq{L0_cost}. We will now illustrate how it may be applied to Adaptive Matching Pursuit~\citep{Perrinet03,Perrinet04} and measure its role on the emergence of edge-like simple cell receptive fields.%
\section{Methods}%
\label{sec:ssc}%
\subsection{Matching Pursuit and Adaptive Matching Pursuit}%
\label{sec:amp}%
Let's first define Adaptive Matching Pursuit. We saw that optimizing the efficiency by minimizing the L$_0$ norm cost leads to a combinatorial search with regard to the dimension of the dictionary. In practice, it means that for a given dictionary, finding the best sparse vector according to minimizing $\mathcal{C}_0( \coef | \image , \dico) $ (see~\seeEq{L0_cost}) is hard and thus that learning an adapted dictionary is difficult. As proposed in~\citep{Perrinet02sparse}, we may solve this problem using a greedy approach. In general, a greedy approach is applied when finding the best combination of elements is difficult to solve globally: A simpler solution is to solve the problem progressively, one element at a time.%

Applied to~\seeEq{L0_cost}, it corresponds to first choosing the single element $a_i \dico_i $ that best fits the image. From the definition of the LGM, we know that for a given signal $\image$, the probability $P( \{a_i\} | \image, \dico )$ corresponding to a \emph{single} source $ a_i \dico_{i}$ for any $i$ is maximal for the dictionary element $i^\ast$ with maximal correlation coefficient:%
\begin{equation}%
i^\ast = \mbox{ArgMax}_i (\rho_i) \mbox{, with } \rho_i = <\frac{\image}{\| \image \|} , \frac{ \dico_i}{\| \dico_i\|} >%
\label{eq:coco}%
\end{equation}%
This formulation is slightly different from Eq.~21 in \citep{Olshausen98}. It should be noted that $\rho_i$ is the $L$-dimensional cosine ($L$ is the dimension of the input space) and that its absolute value is therefore bounded by 1. The value of $\mbox{ArcCos}(\rho_i)$ would therefore give the angle of $\image$ with the pattern $\dico_i$ and in particular, the angle (modulo $2\pi$) would be equal to zero if and only if $\rho_i=1$ (full correlation), $\pi$ if and only if $\rho_i=-1$ (full anti-correlation) and $\pm\pi/2$ if $\rho_i=0$ (both vectors are orthogonal, there is no correlation). The associated coefficient is the scalar projection:%
\begin{equation}%
a_{i^\ast} = <\image , \frac{ \dico_{i^\ast}}{\| \dico_{i^\ast}\|^2} >%
\label{eq:proj}%
\end{equation}%

Second, knowing this choice, the image can be decomposed in%
\begin{equation}
\image = a_{i^\ast} \dico_{i^\ast} + \bf{R} 
\label{eq:mp0}
\end{equation}
where $\bf{R}$ is the residual image. We then repeat this 2-step process on the residual (that is, with $\image \la \bf{R}$) until some stopping criterion is met. 

Hence, we have a sequential algorithm which permits to reconstruct the signal using the list of choices and that we called Sparse Spike Coding~\citep{Perrinet02sparse}. The coding part of the algorithm produces a sparse representation vector $\coef$ for any input image: Its L$_0$ norm is the number of active neurons. Note that the norm of the filters have no influence in this algorithm on the choice function nor on the cost. For simplicity and without loss of generality, we will thereafter set the norm of the filters to $1$: $\forall i, \| A_i \| =1$. It is equivalent to Matching Pursuit (MP) algorithm~\citep{Mallat93} and we have proven previously that this yields an efficient algorithm for representing natural images. Using MP in the SHL scheme defined above (see~\seeSec{shl}) defines Adaptive Matching Pursuit (AMP)~\citep{Perrinet03,Perrinet04} and is similar to other strategies such as~\citep{Smith06, Rehn07}. This class of SHL algorithms offers a non-parametric solution to the emergence of simple cell receptive fields, but compared to \sparsenet , the results often appear to be qualitatively non-homogeneous. Moreover, the heuristic used in \sparsenet\ for the homeostasis may not be used directly since in MP the choice is independent to the norm of the filter. The coding algorithm's efficiency may be improved using Optimized Orthogonal MP~\citep{Rebollo-Neira02} and be integrated in a SHL scheme~\citep{Rehn07}. However, this optimization is separate with the problem that we try to tackle here by optimizing the representation at the learning time scale. Thus, we will now study how we may use  cooperative homeostasis in order to optimize the overall coding efficiency of the dictionary learnt by AMP.
\subsection{Competition-Optimized Matching Pursuit (COMP)}%
In fact, we may now include cooperative homeostasis into AMP. At the coding level, it is important to note that if we simply equalize the sparse output of the MP algorithm, transformed coefficients will indeed be uniformly distributed but the sequence of chosen filters will not be changed. However, the MP algorithm is non-linear and the choice of an element at one step may influence the rest of the choices. This sequence is therefore crucial for the representation efficiency. In order to optimize the competition of the choice step, we may instead choose at every matching step the item in the dictionary corresponding to the most significant value computed thanks to the cooperative homeostasis (see~\seeFig{laughlin}-Right). In practice, it means that we select the best match in the vector corresponding to the transformed coefficients $\bf{z}$, that is, in the vector of the residual coefficients weighted by the non-linearities defined by~\seeEq{laughlin}. This scheme thus extends the MP algorithm which we used previously by linking it to a statistical model which optimally tunes the ArgMax operator in the matching step: Over natural images, for any given dictionary ---and thus independently to the selectivity of the different items from the dictionary--- the choice of a neuron is statistically equally probable. Thanks to cooperative homeostasis, the efficiency of every match in MP is thus maximized, hence the name of Competition-Optimized Matching Pursuit (COMP).

Let's now explicitly describe the COMP coding algorithm step by step. Initially, given the signal $\image$, we set up for all $i$ an internal activity vector $\bar{\coef}$ as the linear correlation using \seeEq{proj}. The output sparse vector is set initially to a zero vector: $\coef={\mathbf 0}$. Using the internal activity $\bar{\coef}$, the neural population will evolve dynamically in an event-based manner by repeating the two following steps. First, the ``Matching'' step is defined by choosing the address with the most significant activity:%
\begin{equation}%
i^\ast = \mbox{ArgMax}_{i} [z_i( \bar{a}_i )]%
\label{eq:mp1}%
\end{equation}%
Then, we set the winning sparse coefficient at address $i^\ast$ with $a_{i^\ast} \la \bar{a}_{i^\ast}$. In the second ``Pursuit'' step, as in MP, the information is fed-back to correlated dictionary elements by:%
\begin{equation}%
\bar{a}_i \la \bar{a}_i - a_{i^\ast} <\dico_{i^\ast} , \dico_i >%
\label{eq:mp3}%
\end{equation}%
Note that after the update, the winning internal activity is zero: $\bar{a}_{i^\ast} = 0$ and that, as in MP, a neuron is selected at most once. Physiologically, as previously described, the pursuit step could be implemented by a lateral, correlation-based inhibition. The algorithm is iterated with~\seeEq{mp1} until some stopping criteria is reached, such as when the residual error energy is below the noise level $\sigma_n^2$. As in MP, since the residual is orthogonal to $ \dico_{i^\ast}$, the residual error energy $E = \| \image \|^2$ may be easily updated at every step as:%
\begin{equation}%
E \la E- a_{i^\ast}^2%
\label{eq:mpe}%
\end{equation}%
COMP transforms the image $\image$ into the sparse vector $\coef$ at any precision $\sqrt{E}$. As in MP, the image may be reconstructed using: $\bar{\image} = \sum_{i} a_{i} \dico_{i}$, which thus gives a solution for~\seeEq{lgm}. COMP differs from MP only by the ``Matching'' step and shares many properties with MP, such as the monotonous decrease of the error (see~\seeEq{mpe}) or the exponential convergence of the coding. However, the decrease of E will always be faster in MP than in COMP from the constraint in the matching step.%

Yet, for a given dictionary, we do not know a priori the functions $z_i$ since they depend on the computation of the sparse coefficients. In practice, the $z_i$ functions are initialized for all neurons to similar arbitrary cumulative distribution functions (COMP is then equivalent to the MP algorithm since choices are not affected). Since we have at most one sparse value $a_i$ per neuron, the cumulative histogram function for each neuron for one coding sweep is $P(a_i \leq \bar{a}_i) = \delta(a_i \leq \bar{a}_i)$ where variable $\bar{a}_i$ is the observed coefficient to be transformed and $\delta$ is the Dirac measure: $\delta(B)=1$ if the boolean variable $B$ is true and $0$ otherwise.  We evaluate~\seeEq{laughlin} after the end of every coding using an online stochastic algorithm, $\forall i, \forall \bar{a}_i$:
\begin{equation}%
z_i( \bar{a}_i ) \la (1- \eta_h ) z_i( \bar{a}_i ) + \eta_h \delta(a_i \leq \bar{a}_i)%
\label{eq:learn_homeo}%
\end{equation}%
where $\eta_h$ is the homeostatic learning rate. Note that this corresponds to the empirical estimation and assumes that coefficients are stationary on a time scale of $\frac{1}{\eta_{h}}$ learning steps. The time scale of homeostasis should therefore in general be less than the time scale of learning. Moreover, due to the exponential convergence of MP, for any set of components, the $z_i$ functions converge to the correct non-linear functions as defined by~\seeEq{laughlin}.%
\subsection{Adaptive Sparse Spike Coding (aSSC)}%
\label{sec:summary}
We may finally apply COMP to Sparse Hebbian Learning (see~\seeSec{shl}). Since the efficiency is inspired by the spiking nature of neural representations, we call this algorithm adaptive Sparse Spike Coding (aSSC). From the definition of COMP, we know that whatever the dictionary, the competition between filters will be fair thanks to the cooperative homeostasis. We add no other homeostatic regulation. We normalize filters' energy since it is a free parameter in~\seeEq{coco}. 

In summary, the whole learning algorithm is given by the following nested loops in pseudo-code:%
\begin{enumerate}%
\item Initialize the point non-linear gain functions $z_i$ to similar cumulative distribution functions and the components $\dico_i$ to random points on the unit $L$-dimensional sphere,%
\item repeat until learning converged:%
\begin{enumerate}%
\item draw a signal $\image$ from the database, its energy is $E = \| \image \|^2$,%
\item set sparse vector $\coef$ to zero, initialize $\bar{a}_i=<\image, \dico_i >$ for all $i$,
\item while the residual energy $E$ is above a given threshold do:
\begin{enumerate}
\item select the best match: $i^\ast = \mbox{ArgMax}_{i} [z_i( \bar{a}_i )]$,
\item set the sparse coefficient: $a_{i^\ast} = \bar{a}_{i^\ast}$,
\item update residual coefficients: $\forall i, \bar{a}_i \la \bar{a}_i - a_{i^\ast} <\dico_{i^\ast} , \dico_i > $,
\item update energy: $E \la E - a_{i^\ast}^2 $.
\end{enumerate}
\item when we have the sparse representation vector $\coef$, apply $\forall i$:
\begin{enumerate}
\item modify dictionary: $\dico_{i} \la \dico_{i} + \eta a_{i} (\image - \dico\coef)$,
\item normalize dictionary: $\dico_{i} \la \dico_{i} / \| \dico_{i}\|$,
\item update homeostasis functions: $z_i( \cdot ) \la (1- \eta_h ) z_i( \cdot ) + \eta_h \delta( a_i \leq \cdot)$.
\end{enumerate}
\end{enumerate}
\end{enumerate}
\section{Results on natural images}%
\label{sec:results}%
The aSSC algorithm differs from the \sparsenet\ algorithm by the MP sparse coding algorithm and by the cooperative homeostasis. Using natural images, we evaluate the relative contribution of these different mechanisms to the representation efficiency.%
\subsection{Receptive field formation}
\label{sec:results_maps}%
\begin{figure}
\centerline{%
\includegraphics[width=\textwidth]{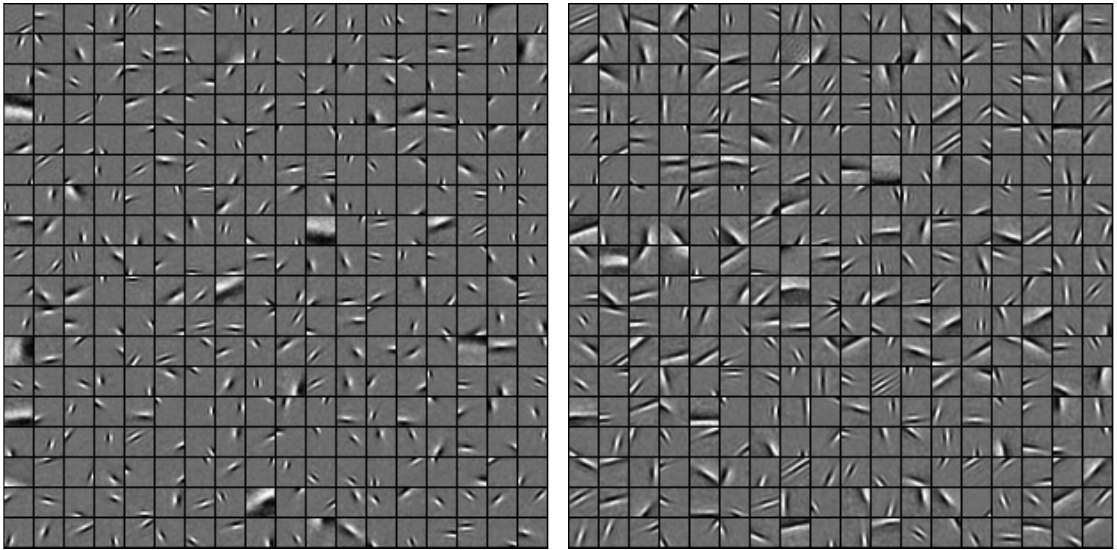}%
}%
\caption{\textbf{Comparison of the dictionaries obtained with {\bf\sc SparseNet} and aSSC.} We show the results of Sparse Hebbian Learning using two different sparse coding algorithms at convergence (20000 learning steps): \textit{(Left)} conjugate gradient function (CGF) method as used in {\sc SparseNet}~\citep{Olshausen98} with \textit{(Right)} COMP as used in aSSC. Filters of the same size as the image patches are presented in a matrix (separated by a black border). Note that their position in the matrix is arbitrary as in ICA.%
}%
\label{fig:filters}%
\end{figure}
We first compare the dictionaries of filters obtained by both methods. We use a similar context and architecture as the experiments described in~\citep{Olshausen98} and specifically the same database of image patches as the \sparsenet\ algorithm. These images are static, grayscale and whitened according to the same parameters to allow a one-to-one comparison of both algorithms. Here, we show the results for $16\times16$ image patches (so that $L=256$) and the learning of $M=324$ filters which are replicated as ON and OFF filters. Assuming this symmetry in the aSSC algorithm, we use the absolute value of the coefficient in~\seeEq{mp1} and~\seeEq{learn_homeo}\footnote{That is, following~\seeSec{summary}, step 2-c-i becomes $i^\ast = \mbox{ArgMax}_{i} [z_i( |\bar{a}_i| )]$, and step 2-d-iii is changed to $ z_i( \cdot ) \la (1- \eta_h ) z_i( \cdot ) + \eta_h \delta( |a_i| \leq \cdot)$.}, the rest of the algorithm being identical. Results replicate the original results of~\citet{Olshausen98} and are comparable for both methods: Dictionaries consist of edge-like filters similarly to the receptive fields of simple cells in the primary visual cortex (see~\seeFig{filters}). Studying the evolution of receptive fields during learning shows that they first represent any salient feature (such as sharp corners or edges), because these correspond to larger Lipschitz coefficients~\citep{Perrinet03ieee}. If a receptive field contains multiple singularities, only the most salient remains later on during learning: Due to the competition between filters, the algorithm eliminates features that are duplicated in the dictionary. Filters which already converged to independent components will be selected sparsely and with high associated coefficients, but inducing a slower learning since corresponding error is small (see~\seeEq{learn}). We observe for both algorithms that when considering very long learning times, the solution is not fixed and edges may slowly drift from one orientation to another while global efficiency remains stable. This is due to the fact that there are many solutions to the same problem (note, for instance, that solutions are invariant up to a permutation of neurons' addresses). It is possible to decrease these degrees of freedom by including for instance topological links between filters~\citep{Bednar04}. Qualitatively, the main difference between both results is that filters produced by aSSC look more diverse and broad (so that they often overlap), while the filters produced by \sparsenet\ are more localized and thin.%

We also perform robustness experiments to determine the range of learning parameters for which these algorithms converged. One advantage of aSSC is that it is based on a non-parametric sparse coding and a non-parametric homeostasis rule and is entirely described by $2$ structural parameters ($L$ and $M$) and $2$ learning parameters ($\eta$ and $\eta_h$) while parameterization of the prior and of the homeostasis for \sparsenet\ requires $5$ more parameters to adjust ($3$ for the prior, $2$ for the homeostasis). By observing at convergence the probability distribution function of selected filters, homeostasis in aSSC converges for a wide range of $\eta_h$ values (see~\seeEq{learn_homeo}). Furthermore, we observe that at convergence, the $z_i$ functions become very similar (see dotted lines in~\seeFig{laughlin}-Right) and that homeostasis does not favor the selection of any particular neuron as strongly as at the beginning of the learning. Therefore, thanks to the homeostasis, equilibrium is reached when the dictionary homogeneously represents different features in natural images, that is, when filters have similar selectivities. Finally, we observe the counter-intuitive result that non-linearities implementing cooperative homeostasis are important for the coding only \emph{during} the learning period but that it may be ignored for the coding after convergence since at this point non-linearities are the same for all neurons.%

Both dictionaries appear to be qualitatively different and for instance parameters of the emerging edges (frequency, length, width) are distributed differently. In fact, it seems that rather than the shape of each dictionary element taken individually, it is their distribution in image space that yields different efficiencies. Such an analysis of the filters' shape distribution was performed quantitatively for \sparsenet\ in~\citep{Lewicki00}. The filters were fitted by Gabor functions~\citep{Jones87}. A recent study compares the distribution of fitted Gabor functions' parameters between the model and receptive fields obtained from neurophysiological experiments conducted in primary visual cortex of macaques~\citep{Rehn07}. It has shown that their SHL model based on Optimized Orthogonal MP better matches to physiological observations than \sparsenet . However, there is no theoretical basis for the fact that receptive fields' shape should be well fitted by Gabor functions~\citep{Saito01} and the variety of shapes observed in biological systems may for instance reflect adaptive regulation mechanisms when reaching different optimal sparseness levels~\citep{Assisi07}. Moreover, even though this type of quantitative method is certainly necessary, it is not sufficient to understand the role of each individual mechanism in the emergence of edge-like receptive fields. To asses the relative role of coding and homeostasis in SHL, we rather compare these different dictionaries quantitatively in terms of representation efficiency.%
\subsection{Coding efficiency in SHL}%
\label{sec:results_efficiency}%
\begin{figure}
\begin{center}%
\includegraphics[width=\textwidth]{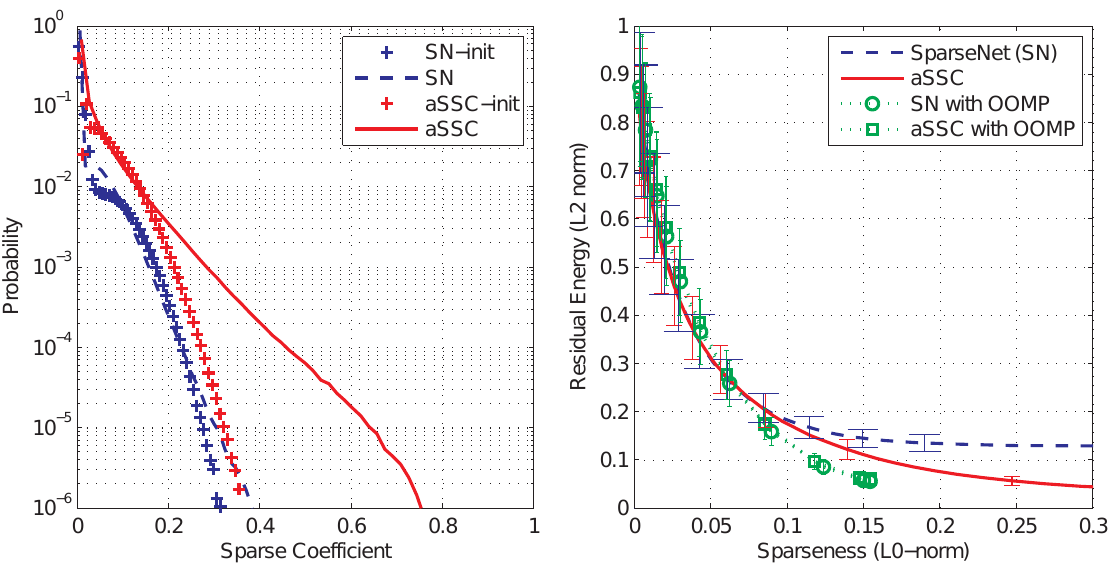}%
\end{center}%
\caption{\textbf{Coding efficiency of \sparsenet\ versus aSSC.} We evaluate the quality of both learning schemes by comparing coding efficiency of their respective coding algorithms, that is CGF and COMP, with the respective dictionary that was learnt (see~\seeFig{filters}). {\it (Left)} We show the probability distribution function of sparse coefficients obtained by both methods with random dictionaries (respectively 'SN-init' and 'aSSC-init') and with the dictionaries obtained after convergence of respective learning schemes (respectively 'SN' and 'aSSC'). At convergence, sparse coefficients are more sparsely distributed than initially, with more kurtotic probability distribution functions for aSSC in both cases.{\it (Right)} We plot the average residual error (L$_2$ norm) as a function of the relative number of active (non-zero) coefficients. This provides a measure of the coding efficiency for each dictionary over the set of image patches (error bars are scaled to one standard deviation).  The L$_0$ norm is equal to the coding step in COMP. Best results are those providing a lower error for a given sparsity (better compression) or a lower sparseness for the same error (Occam's razor). We observe similar coding results in aSSC despite its non-parametric definition. This result is also true when using the two different dictionaries with the same OOMP sparse coding algorithm: The dictionaries still have similar coding efficiencies.  
}%
\label{fig:efficiency}%
\end{figure}
To address this issue, we first compare the quality of both dictionaries (from \sparsenet\ and aSSC) by computing the mean efficiency of their respective \emph{coding} algorithms (respectively~CGF and COMP). Using $10^5$ image patches drawn from the natural image database, we perform the progressive coding of each image using both sparse coding methods. When plotting the probability distribution function of the sparse coefficients, one observes that distributions fit well the bivariate model introduced in~\citep{Olshausen00nips} where a sub-set of the coefficients are null (see~\seeFig{efficiency}-Left). Log-probability distributions of non-zero coefficients is quadratic with the initial random dictionaries. At convergence, non-zero coefficients fit well to a Laplacian probability distribution function. Measuring mean kurtosis of resulting sparse vectors proves to be very sensitive and a poor indicator of global efficiency, in particular at the beginning of the coding, when many coefficients are still strictly zero. In general, COMP provides a sparser final distribution. Dually, plotting the decrease of the sorted coefficients as a function of their rank shows that coefficients for COMP are first higher and then decrease more quickly, due to the link between the $z_i$ functions and the function of sorted coefficients (see~\seeEq{laughlin}). As a consequence,  a Laplacian bivariate model for the distribution of sparse coefficient emerge from the statistics of natural images. The advantage of aSSC is that this emergence is not dependent of a parametric model of the prior.

In a second analysis, we compare the efficiency of both methods while varying the number of active coefficients (the L$_0$ norm). We perform this in COMP by simply measuring the residual error (L$_2$ norm) with respect to the coding step. To compare this method with the conjugate gradient method, we use a 2-pass sparse coding: A first pass identifies best neurons for a fixed number of active coefficients, while a second pass optimizes the coefficients for this set of ``active'' vectors. This method was also used in~\citep{Rehn07} and proved to be fair when comparing both algorithms. We observe in a robust manner that the greedy solution to the hard problem (that is, COMP) is as efficient as conjugate gradient as used in \sparsenet\ (see~\seeFig{efficiency}, Right). We also observe that aSSC is also slightly more efficient for the cost defined in~\seeEq{sparse_cost}, a result which may reflect the fact that the L$_0$ norm defines a stronger sparseness constraint than the convex cost. Moreover, we compare the coding efficiency of both dictionaries using Optimized Orthogonal MP. Results show that OOMP provides a slight coding improvement, but also confirms that both dictionaries are of similar coding efficiency, independently of their respective coding algorithm. 

These results prove that, without the need of a parameterization of the prior, coding in aSSC is as efficiency than \sparsenet . In addition, there are a number of other advantages offered by this approach. First, COMP simply uses a feed-forward pass with lateral interactions, while conjugate gradient is implemented as the fixed point of a recurrent network (see Figure 13.2 from~\citep{Olshausen02}). Moreover, we have already seen that aSSC is a non-parametric method which is controlled by fewer parameters. Therefore, applying a ``higher-level'' Occam razor confirms that for a similar overall coding efficiency, aSSC is better since it is of lower \emph{structural} complexity\footnote{A quantitative measure of the structural complexity for the different methods is given by the minimal length of a code that would implement them, this length being defined as the number of characters of the code implementing the algorithm. It would therefore depend on the machine on which it is implemented, and there is, of course, a clear advantage of aSSC on parallel architectures.}. Finally, in \sparsenet\ and in algorithms defined in~\citep{Lewicki00,Smith06,Rehn07}, representation is analog without explicitly defining a quantization.  This is not the case in the aSSC algorithm where cooperative homeostasis introduces a regularity in the distribution of sparse coefficients. %
\subsection{Role of homeostasis in representation efficiency}%
\label{sec:results_homeo}%
\begin{figure}
\begin{center}%
\includegraphics[width=\textwidth]{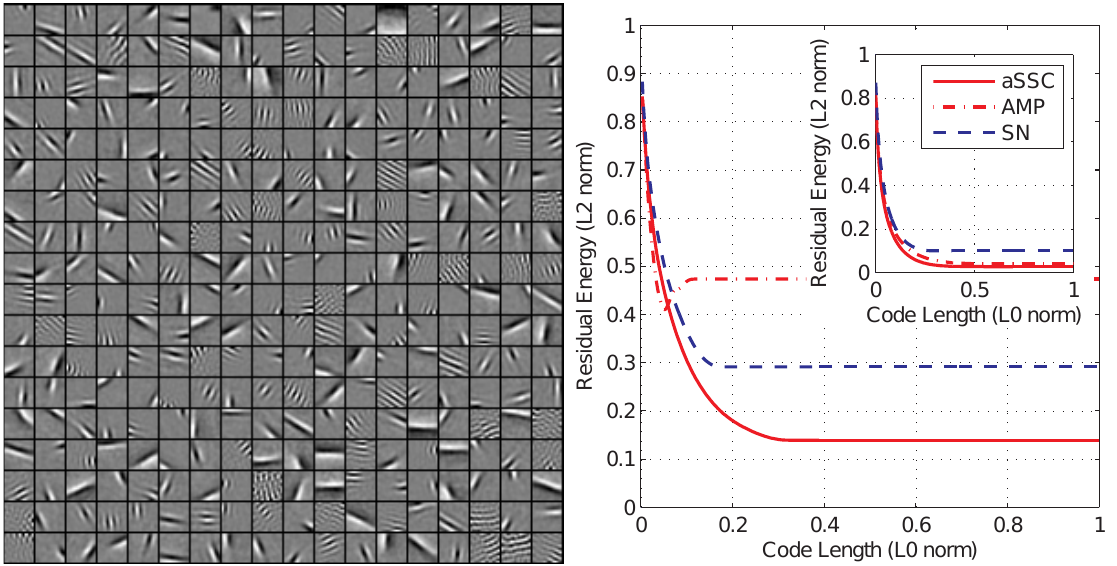}%
\end{center}%
\caption{\textbf{Cooperative homeostasis implements efficient quantization.} {\it (Left)} When switching off the cooperative homeostasis during learning, the corresponding Sparse Hebbian Learning algorithm, Adaptive Matching Pursuit (AMP), converges to a set of filters that contains some less localized filters and some high-frequency Gabor functions that correspond to more ``textural'' features~\citep{Perrinet03}. One may wonder if these filters are inefficient and capturing noise or if they rather correspond to independent features of natural images in the LGM model. {\it (Right, Inset)} In fact, when plotting residual energy as a function of L$_0$ norm sparseness with the MP algorithm (as plotted in~\seeFig{efficiency}, Right), the AMP dictionary gives a slightly worse result than aSSC. {\it (Right)} Moreover, one should consider representation efficiency as the overall coding and decoding algorithm. We compare the efficiency for these dictionaries thanks to same coding method (SSC) and the same decoding method (using rank quantized coefficients). Representation length for this decoding method is proportional to the L$_0$ norm with $\lambda=\frac{\log(M)}{L} \approx 0.032$ bits per coefficient and per pixel as defined in~\seeEq{L0_cost}. We observe that the dictionary obtained by aSSC is more efficient than the one obtained by AMP while the dictionary obtained with  \sparsenet\ (SN) gives an intermediate result thanks to the geometric homeostasis: Introducing cooperative homeostasis globally improves neural representation. %
}%
\label{fig:homeo}%
\end{figure}
In the context of an information channel such as implemented by a neural assembly, one should rather use the coefficients that could be decoded from the neural signal in order to define the reconstruction cost (see~\seeFig{laughlin}, Left). As was described in \seeSec{lgm}, knowing a dictionary $\dico$, it is indeed more correct to consider the overall average coding and decoding cost over image patches $\mathcal{C}(\hat{\coef} | \image, \dico)$ (see~\seeEq{efficiency_cost}), where $\hat{\coef}$ corresponds to the analog vector of coefficients inferred from the neural representation. The overall transmission error may be described as the sum of the reconstruction and the quantization error. This last error will increase both with inter-trial variability but also with the non homogeneity of the represented features. It is however difficult to evaluate a decoding scheme in most sparse coding algorithms since this problem is generally not addressed. Our objective when defining $\mathcal{C}_0(\hat{\coef} | \image, \dico)$ (see~\seeEq{L0_cost}) was to define sparseness as it may be represented by spiking neural representations. Using a decoding algorithm on such a representation will help us to quantify overall coding efficiency. 

An effective decoding algorithm is to estimate the analog values of the sparse vector (and thus reconstruct the signal) from the order of neurons' activation in the sparse vector~\citep[Section 2.2]{Perrinet06}. In fact, knowing the address of the fiber $i^0$ corresponding to the maximal value, we may infer that it has been produced by an analog value on the emitter side in the highest quantile of the probability distribution function of $a_{i^0}$. We may therefore decode the corresponding value with the best estimate which is given as the average maximum sparse coefficient for this neuron by inverting $z_{i^0}$ (see~\seeEq{laughlin})\footnote{Mathematically, the $z_i$ are not always \emph{strictly} increasing and we state here that $z_{i}^{-1}(z)$ is defined in a unique way as the average value of the coefficients $a_i$ such that $z_i(a_i)=z$.}: ${a}_{i^0} = z_{i^0}^{-1}(1)$. This is also true for the following coefficients. We write as $\frac{r}{M}$ the relative rank of the $r^{th}$ and $o$ the order function which gives the address of the winning neuron at rank $r$. Since $z_{o(r)}= 1- \frac{r}{M} = z_{o(r)}(a_{o(r)})$, we can reconstruct the corresponding value as%
\begin{equation}%
\hat{a}_{o(r)} = z_{o(r)}^{-1}(1- \frac{r}{M} )%
\label{eq:mod}%
\end{equation}%
Physiologically, \seeEq{mod} could be implemented using interneurons which would ``count'' the number of received spikes and by modulating efficiency of synaptic events on receiver efferent neurons ---for instance with shunting inhibition~\citep{Delorme03}. Recent findings show that this type of code may be used in cortical in vitro recurrent networks \citep{Shahaf08}. This corresponds to a generalized rank coding scheme. However this quantization does not require that neural information explicitly carries rank information. In fact, this scheme is rather general and is analogous to scalar quantization using the modulation function $z_i^{-1}$ as a Look-Up-Table. It is very likely that fine temporal information such as inter-spike intervals also play a role in neural information transmission. As in other decoding schemes, the quantization error directly depends on the variability of the modulation functions across trials~\citep{Perrinet03ieee}. This scheme thus rather shows a representative behavior for the retrieval of information from spiking neural activity. 

To evaluate the specific role of cooperative homeostasis, we compare previous dictionaries (see~\seeFig{filters}) with the one obtained by Adaptive Matching Pursuit (AMP). In fact, \sparsenet\ and aSSC differ at the level of the homeostasis but also for the sparse coding. The only difference between aSSC and AMP is the introduction of cooperative homeostasis. To obtain the solution to AMP, we use the same sparse coding algorithm but switch off the cooperative homeostasis during learning ($\eta_h = 0$ in~\seeEq{learn_homeo}). We observe at convergence that the dictionary corresponds qualitatively  to features which are different from aSSC and \sparsenet\  (see~\seeFig{homeo}, Left). In particular, we observe the emergence of Gabor functions with broader width which better match textures. These filters correspond to lower Lipschitz coefficients~\citep{Perrinet03ieee}, and because of their lower saliency, these textural filters are more likely to be selected with lower correlation coefficients. They fit more to Fourier filters that are obtained using Principal Component Analysis~\citep{Fyfe95} and are still optimal to code arbitrary image patches such as noise~\citep{Zhaoping06}. When we plot L$_2$ norm with respect to L$_0$ norm for the different dictionaries with the same MP coding algorithm averaged over a set of $10^5$ image patches from natural scenes (see~\seeFig{homeo}, Right Inset), the resulting dictionary from AMP is less efficient than those obtained with aSSC and \sparsenet . This is not an expected behavior since COMP is more constrained than MP (MP is the ``greediest'' solution) and using both methods with a similar dictionary would necessarily give an advantage to MP: the AMP thus reached a local minima of the coding cost. To understand why, recall that in the aSSC algorithm, the cooperative homeostasis constraint, by its definition in~\seeEq{laughlin}, plays the role of a gain control and that the point non-linearity from~\seeEq{mp1} ensures that all filters are selected equally. Compared to AMP, textured elements are ``boosted'' during learning relative to a more generic salient edge component and are thus more likely to evolve (see~\seeFig{laughlin}, Right). This explains why they would end up being less probable and that at convergence there are no textured filters in the dictionary obtained with aSSC.

Finally, we test quantitatively representation efficiency of these different dictionaries  with the same quantization scheme. At the decoding level, we compute in all cases the modulation functions as defined in~\seeEq{mod} on a set of $10^5$ image patches from natural scenes. Since addresses' choices may be generated by any of the $M$ neurons, the representation cost is defined as $\lambda=\log(M)$ bits per chosen address (see~\seeEq{L0_cost}). Then, when using the quantization (see~\seeEq{mod}), the AMP approach displays a larger variability reflecting the lack of homogeneity of the features represented by the dictionary: There is a much larger reconstruction error and a slower decrease of error's energy (see~\seeFig{homeo}, Right). The aSSC on the contrary is adapted to quantization thanks to the cooperative homeostasis and consequently yields a more regular decrease of coefficients as a function of rank, that is, a lower quantization error. The dictionary obtained with the \sparsenet\  algorithm yields an intermediate result. This shows that the heuristic implementing the homeostasis in this algorithm regulates relatively well the choices of the elements during the learning. It also explains why the three parameters of the homeostasis algorithm had to be properly tuned to fit the dynamics of the heuristics. Results therefore show that homeostasis optimizes the efficiency of the neural representation during learning and that the cooperative homeostasis provides a simple and effective optimization scheme.%
\section{Discussion}%
\label{sec:discussion}%
We have shown in this paper that homeostasis plays an essential role in Sparse Hebbian Learning (SHL)  schemes and thus on our understanding of the emergence of simple cell receptive fields. First, using statistical inference and information theory, we have proposed a quantitative cost for the coding efficiency based on a non-parametric model using the number of active neurons, that is, the L$_0$ norm of the representation vector. This allowed to design a cooperative homeostasis rule based on neurophysiological observations~\citep{Laughlin81}. This rule optimizes the competition between neurons by simply constraining the choice of every selection of an active neuron to be equiprobable. This homeostasis defined a new sparse coding algorithm, COMP, and a new SHL scheme, aSSC. Then, we have confirmed that the aSSC scheme provides an efficient model for the formation of simple cell receptive fields, similarly to other approaches. The sparse coding algorithms in these schemes are variants of conjugate gradient or of Matching Pursuit. They are based on correlation-based inhibition since this is necessary to remove redundancies from the linear representation. This is consistent with the observation that lateral interactions are necessary for the formation of elongated receptive fields~\citep{Bolz89}. With a correct tuning of parameters, all schemes show the emergence of edge-like filters. The specific coding algorithm used to obtain this sparseness appears to be of secondary importance as long as it is adapted to the data and yields sufficiently efficient sparse representation vectors. However, resulting dictionaries vary qualitatively among these schemes and it was unclear which algorithm is the most efficient and what was the individual role of the different mechanisms that constitute SHL schemes. At the learning level, we have shown that the homeostasis mechanism had a great influence on the qualitative distribution of learned filters. In particular, using the comparison of coding and decoding efficiency of aSSC with and without this specific homeostasis, we have proven that cooperative homeostasis optimized overall representation efficiency. This efficiency is comparable with that of \sparsenet\ , but with the advantage that our unsupervised learning model is non-parametric and does not need to be properly tuned. 

This work might be advantageously applied to signal processing problems. First, we saw that optimizing the representation cost maximizes the independence between features and is related to the goal of ICA. Since we have built a solution to the LGM inverse problem that is more efficient than standard methods such as the \sparsenet\ algorithm, it is thus a good candidate solution to Blind Source Separation problems. Second, at the coding level, we optimized in the COMP algorithm the efficiency of Matching Pursuit by including an adaptive cooperative homeostasis mechanism. We proved that for a given compression level, image patches are more efficiently coded than in the Matching Pursuit algorithm. Since we have shown previously that MP compares favorably with compression methods such as JPEG with a fixed log-Gabor filter dictionary~\citep{Fischer07}, we can predict that COMP should provide promising results for image representation. An advantage over other sparse coding schemes is that it provides a progressive dynamical result while the conjugate gradient method has to be recomputed for every different number of coefficients. The most relevant information is propagated first and progressive reconstruction may be interrupted at any time. Finally, a main advantage of this type of neuromorphic algorithm is that it uses a simple set of operations: computing the correlation, applying the point non-linearity from a Look-Up Table, choosing the $\mbox{ArgMax}$, doing a subtraction, retrieving a value from a Look-Up-Table. In particular, the complexity of these operations, such as the $\mbox{ArgMax}$ operator, would in theory not depend on the dimension of the system in parallel machines and the transfer of this technology to neuromorphic hardware such as aVLSIs~\citep{Schemmel06,Bruderle09} will provide a supra-linear gain of performance.%

In this paper, we focused on transient input signals and of relatively abstract neurons. This choice was made to highlight the powerful function of the parallel and temporal competition between neurons in contrast to traditional analog and sequential strategies using analog spike frequency representations. This strategy allowed to compare the proposed learning scheme with state-of-the-art algorithms. One obvious extension to the algorithm is to implement learning with more realistic inputs. In fact, sparseness in image patches is only local while it is also spatial and temporal in whole-field natural scenes: For instance, it is highly probable in whole natural images that large parts of the space ---such as the sky--- are flat and contain no information. Our results should be thus taken as a lower bound for the efficiency of aSSC in natural scenes. This also suggests the extension to representations with some built-in invariances, such as translation and scaling. A gaussian pyramid, for instance, provides a multi-scale representation where the set of learned filters would  become a dictionary of mother wavelets~\citep[Section 3.3.4]{Perrinet06}. Such an extension leads to a fundamental question: How does representation efficiency evolves with the number $M$ of elements in the dictionary, that is, with the complexity of the representation? In fact, when increasing the over-completeness in aSSC, one observes the emergence of different classes of edge filters: at first different positions, then different orientations of edges, followed by different frequencies and so on and so forth. This specific order indicates the existence of an underlying hierarchy for the synthesis of natural scenes. This hierarchy seems to correspond to the level of importance of the different transformations that are learned by the system, respectively translation, rotation and scaling. Exploring the efficiency results for different dimensions of the dictionary in aSSC will thus give a quantitative evaluation of the optimal complexity of the model needed to describe images in terms of a trade-off between accuracy and generality. But it may also provide a model for the clustering of the low-level visual system into different areas, such as the emergence of position-independent representations in the ventral visual pathway versus motion-selective neurons in the dorsal visual pathway.%
\subsubsection*{Acknowledgments}
This work was supported by a grant from the French Research Council (ANR ``NatStats'') and by EC IP project FP6-015879, ``FACETS''. The author thanks the team at the Redwood Neuroscience Institute for stimulating discussions and particularly Jeff Hawkins, Bruno Olshausen, Fritz Sommer, Tony Bell, Dileep George, Kilian Koepsell and Matthias Bethge. Special thanks to Jo Hausmann, Guillaume Masson, Nicole Voges, Willie Smit and Artemis Kosta for essential comments on this work.  We thank the anonymous referees for their helpful comments on the manuscript. Special thanks to Bruno Olshausen, Laura Rebollo-Neira, Gabriel Peyr\'e, Martin Rehn  and Fritz Sommer for providing the source code for their experiments.%


\begin{thebibliography}{51}
\expandafter\ifx\csname natexlab\endcsname\relax\def\natexlab#1{#1}\fi
\expandafter\ifx\csname url\endcsname\relax
  \def\url#1{{\tt #1}}\fi
\expandafter\ifx\csname urlprefix\endcsname\relax\def\urlprefix{}\fi

\bibitem[{Akaike(1974)}]{Akaike74}
Akaike, H. (1974).
\newblock A new look at the statistical model identification.
\newblock {\em I{EEE} Transactions on Automatic Control\/}, {\em 19\/},
  716--23.

\bibitem[{Assisi et~al.(2007)Assisi, Stopfer, Laurent, \& Bazhenov}]{Assisi07}
Assisi, C., Stopfer, M., Laurent, G., \& Bazhenov, M. (2007).
\newblock Adaptive regulation of sparseness by feedforward inhibition.
\newblock {\em Nature {N}euroscience\/}, {\em 10\/}(9), 1176--84.

\bibitem[{Atick(1992)}]{Atick92}
Atick, J.~J. (1992).
\newblock Could information theory provide an ecological theory of sensory
  processing?
\newblock {\em Network: {C}omputation in {N}eural {S}ystems\/}, {\em 3\/}(2),
  213--52.

\bibitem[{Barlow(2001)}]{Barlow01}
Barlow, H.~B. (2001).
\newblock Redundancy reduction revisited.
\newblock {\em Network: {C}omputation in {N}eural {S}ystems\/}, {\em 12\/},
  241---25.

\bibitem[{Baudot et~al.(2004)Baudot, Levy, Monier, Chavane, Ren{\'e}, Huguet,
  Marre, Pananceau, Kopysova, \& Fregnac}]{Baudot04}
Baudot, P., Levy, M., Monier, C., Chavane, F., Ren{\'e}, A., Huguet, N., Marre,
  O., Pananceau, M., Kopysova, I., \& Fregnac, Y. (2004).
\newblock Time-coding, low noise {V}m attractors, and trial-by-trial spiking
  reproducibility during natural scene viewing in {V}1 cortex.
\newblock In {\em Society for Neuroscience Abstracts: 34th Annual Meeting of
  the Society for Neuroscience, San Diego, USA} (Eds.), (pp. 948--12).

\bibitem[{Bednar et~al.(2004)Bednar, Kelkar, \& Miikkulainen}]{Bednar04}
Bednar, J.~A., Kelkar, A., \& Miikkulainen, R. (2004).
\newblock Scaling self-organizing maps to model large cortical networks.
\newblock {\em Neuroinformatics\/}, {\em 2\/}(3), 275--302.

\bibitem[{Bethge et~al.(2003)Bethge, Rotermund, \& Pawelzik}]{Bethge03}
Bethge, M., Rotermund, D., \& Pawelzik, K. (2003).
\newblock Second order phase transition in neural rate coding: Binary encoding
  is optimal for rapid signal transmission.
\newblock {\em Physical Review Letters\/}, {\em 90\/}(8), 088104.

\bibitem[{Bolz \& Gilbert(1989)}]{Bolz89}
Bolz, J., \& Gilbert, C.~D. (1989).
\newblock The role of horizontal connections in generating long receptive
  fields in the cat visual cortex.
\newblock {\em European Journal of Neuroscience\/}, {\em 1\/}(3), 263--8.

\bibitem[{Br{\"u}derle et~al.(2009)Br{\"u}derle, M{\"u}ller, Davison, Muller,
  Schemmel, \& Meier}]{Bruderle09}
Br{\"u}derle, D., M{\"u}ller, E., Davison, A., Muller, E., Schemmel, J., \&
  Meier, K. (2009).
\newblock Establishing a novel modeling tool: a python-based interface for a
  neuromorphic hardware system.
\newblock {\em Frontiers in Neuroinformatics\/}, {\em 3\/}, 17.

\bibitem[{Chapman \& Stryker(1992)}]{Chapman92}
Chapman, B., \& Stryker, M.~P. (1992).
\newblock Origin of orientation tuning in the visual cortex.
\newblock {\em Current {O}pinion in {N}eurobiology\/}, {\em 2\/}(4), 498--501.

\bibitem[{Delorme \& Thorpe(2003)}]{Delorme03}
Delorme, A., \& Thorpe, S.~J. (2003).
\newblock Early cortical orientation selectivity: {H}ow fast shunting
  inhibition decodes the order of spike latencies.
\newblock {\em Journal of {C}omputational {N}euroscience\/}, {\em 15\/},
  357--65.

\bibitem[{DeWeese et~al.(2003)DeWeese, Wehr, \& Zador}]{Deweese03}
DeWeese, M.~R., Wehr, M., \& Zador, A.~M. (2003).
\newblock Binary coding in auditory cortex.
\newblock {\em Journal of {N}euroscience\/}, {\em 23\/}(21).

\bibitem[{Doi et~al.(2007)Doi, Balcan, \& Lewicki}]{Doi07}
Doi, E., Balcan, D.~C., \& Lewicki, M.~S. (2007).
\newblock Robust coding over noisy overcomplete channels.
\newblock {\em I{EEE} {T}ransactions in {I}mage {P}rocessing\/}, {\em 16\/}(2),
  442--52.

\bibitem[{Field(1994)}]{Field94}
Field, D.~J. (1994).
\newblock What is the goal of sensory coding?
\newblock {\em Neural {C}omputation\/}, {\em 6\/}(4), 559--601.

\bibitem[{Fischer et~al.(2007)Fischer, Redondo, Perrinet, \&
  Crist{\'o}bal}]{Fischer07}
Fischer, S., Redondo, R., Perrinet, L., \& Crist{\'o}bal, G. (2007).
\newblock Sparse approximation of images inspired from the functional
  architecture of the primary visual areas.
\newblock {\em EURASIP Journal on Advances in Signal Processing\/}, {\em
  2007\/}(1), 122.

\bibitem[{Fyfe \& Baddeley(1995)}]{Fyfe95}
Fyfe, C., \& Baddeley, R.~J. (1995).
\newblock Finding compact and sparse- distributed representations of visual
  images.
\newblock {\em Network: {C}omputation in {N}eural {S}ystems\/}, {\em 6\/},
  333--44.

\bibitem[{Hebb(1949)}]{Hebb49}
Hebb, D.~O. (1949).
\newblock {\em The organization of behavior: {A} neuropsychological theory\/}.
\newblock New York: Wiley.

\bibitem[{Hosoya et~al.(2005)Hosoya, Baccus, \& Meister}]{Hosoya05}
Hosoya, T., Baccus, S.~A., \& Meister, M. (2005).
\newblock Dynamic predictive coding by the retina.
\newblock {\em Nature\/}, {\em 436\/}(7047), 71--7.

\bibitem[{Jones \& Palmer(1987)}]{Jones87}
Jones, J.~P., \& Palmer, L.~A. (1987).
\newblock An evaluation of the two-dimensional gabor filter model of simple
  receptive fields in cat striate cortex.
\newblock {\em Journal of {N}europhysiology\/}, {\em 58\/}(6), 1233--58.

\bibitem[{Laughlin(1981)}]{Laughlin81}
Laughlin, S.~B. (1981).
\newblock A simple coding procedure enhances a neuron's information capacity.
\newblock {\em Zeitung f{\"u}r {N}aturforschung\/}, {\em 9--10\/}(36), 910--2.

\bibitem[{Lee \& Seung(1999)}]{Lee99}
Lee, D.~D., \& Seung, H.~S. (1999).
\newblock Learning the parts of objects by non-negative matrix factorization.
\newblock {\em Nature\/}, {\em 401\/}, 788--91.

\bibitem[{Lee et~al.(2007)Lee, Battle, Raina, \& Ng}]{Lee07}
Lee, H., Battle, A., Raina, R., \& Ng, A. (2007).
\newblock Efficient sparse coding algorithms.
\newblock In B.~Sch\"{o}lkopf, J.~Platt, \& T.~Hoffman (Eds.) {\em Advances in
  Neural Information Processing Systems 19\/}, (pp. 801--808). Cambridge, MA:
  MIT Press.

\bibitem[{Lewicki \& Sejnowski(2000)}]{Lewicki00}
Lewicki, M.~S., \& Sejnowski, T.~J. (2000).
\newblock Learning overcomplete representations.
\newblock {\em Neural {C}omputation\/}, {\em 12\/}(2), 337--65.

\bibitem[{Mallat(1998)}]{Mallat98}
Mallat, S. (1998).
\newblock {\em A wavelet tour of signal processing\/}.
\newblock Academic Press, second edition.

\bibitem[{Mallat \& Zhang(1993)}]{Mallat93}
Mallat, S., \& Zhang, Z. (1993).
\newblock Matching {P}ursuit with time-frequency dictionaries.
\newblock {\em I{EEE} {T}ransactions on {S}ignal {P}rocessing\/}, {\em
  41\/}(12), 3397--3414.

\bibitem[{Nikitin et~al.(2009)Nikitin, Stocks, Morse, \& McDonnell}]{Nikitin09}
Nikitin, A.~P., Stocks, N.~G., Morse, R.~P., \& McDonnell, M.~D. (2009).
\newblock Neural population coding is optimized by discrete tuning curves.
\newblock {\em Physical {R}eview {L}etters\/}, {\em 103\/}(13), 138101.

\bibitem[{Oja(1982)}]{Oja82}
Oja, E. (1982).
\newblock A {S}implified {N}euron {M}odel as a {P}rincipal {C}omponent
  {A}nalyzer.
\newblock {\em Journal of {M}athematical biology\/}, {\em 15\/}, 267--73.

\bibitem[{Olshausen(2002)}]{Olshausen02}
Olshausen, B.~A. (2002).
\newblock Sparse codes and spikes.
\newblock In R.~P.~N. Rao, B.~A. Olshausen, \& M.~S. Lewicki (Eds.) {\em
  Probabilistic {M}odels of the {B}rain: {P}erception and {N}eural
  {F}unction\/}, chap. Sparse Codes and Spikes, (pp. 257--72). MIT Press.

\bibitem[{Olshausen \& Field(1996)}]{Olshausen96a}
Olshausen, B.~A., \& Field, D.~J. (1996).
\newblock Emergence of simple-cell receptive field properties by learning a
  sparse code for natural images.
\newblock {\em Nature\/}, {\em 381\/}(6583), 607--9.

\bibitem[{Olshausen \& Field(1997)}]{Olshausen98}
Olshausen, B.~A., \& Field, D.~J. (1997).
\newblock Sparse coding with an overcomplete basis set: a strategy employed by
  {V}1?
\newblock {\em Vision {R}esearch\/}, {\em 37\/}, 3311--25.

\bibitem[{Olshausen \& Millman(2000)}]{Olshausen00nips}
Olshausen, B.~A., \& Millman, K.~J. (2000).
\newblock Learning sparse codes with a mixture-of-gaussians prior.
\newblock In M.~I. Jordan, M.~J. Kearns, \& S.~A. Solla (Eds.) {\em Advances in
  neural information processing systems\/}, vol.~12, (pp. 887--93). The MIT
  Press, Cambridge, MA.

\bibitem[{Pece(2002)}]{Pece02}
Pece, A. E.~C. (2002).
\newblock The problem of sparse image coding.
\newblock {\em Journal of Mathematical Imaging and Vision\/}, {\em 17\/},
  89--108.

\bibitem[{Perrinet(2004)}]{Perrinet04}
Perrinet, L. (2004).
\newblock Finding {I}ndependent {C}omponents using spikes : a natural result of
  hebbian learning in a sparse spike coding scheme.
\newblock {\em Natural {C}omputing\/}, {\em 3\/}(2), 159--75.

\bibitem[{Perrinet(2007)}]{Perrinet06}
Perrinet, L. (2007).
\newblock Dynamical neural networks: modeling low-level vision at short
  latencies.
\newblock In {\em Topics in Dynamical Neural Networks: From Large Scale Neural
  Networks to Motor Control and Vision\/}, vol. 142 of {\em The European
  Physical Journal (Special Topics)\/}, (pp. 163--225). Springer Verlag (Berlin / Heidelberg).

\bibitem[{Perrinet et~al.(2002)Perrinet, Samuelides, \&
  Thorpe}]{Perrinet02sparse}
Perrinet, L., Samuelides, M., \& Thorpe, S.~J. (2002).
\newblock Sparse spike coding in an asynchronous feed-forward multi-layer
  neural network using {M}atching {P}ursuit.
\newblock {\em Neurocomputing\/}, {\em 57C\/}, 125--34.

\bibitem[{Perrinet et~al.(2003)Perrinet, Samuelides, \& Thorpe}]{Perrinet03}
Perrinet, L., Samuelides, M., \& Thorpe, S.~J. (2003).
\newblock Emergence of filters from natural scenes in a sparse spike coding
  scheme.
\newblock {\em Neurocomputing\/}, {\em 58--60\/}(C), 821--6.

\bibitem[{Perrinet et~al.(2004)Perrinet, Samuelides, \&
  Thorpe}]{Perrinet03ieee}
Perrinet, L., Samuelides, M., \& Thorpe, S.~J. (2004).
\newblock Coding static natural images using spiking event times: do neurons
  cooperate?
\newblock {\em I{EEE} {T}ransactions on {N}eural {N}etworks\/}, {\em 15\/}(5),
  1164--75.
\newblock {S}pecial issue on '{T}emporal {C}oding for {N}eural {I}nformation
  {P}rocessing'.

\bibitem[{Ranzato et~al.(2007)Ranzato, Poultney, Chopra, \& LeCun}]{Ranzato07}
Ranzato, M.~A., Poultney, C.~S., Chopra, S., \& LeCun, Y. (2007).
\newblock Efficient learning of sparse overcomplete representations with an
  energy-based model.
\newblock In B.~Sch\"{o}lkopf, J.~Platt, \& T.~Hoffman (Eds.) {\em Advances in
  neural information processing systems\/}, vol.~19, (pp. 1137--44). Cambridge,
  MA: The MIT Press.

\bibitem[{Rebollo-Neira \& Lowe(2002)}]{Rebollo-Neira02}
Rebollo-Neira, L., \& Lowe, D. (2002).
\newblock Optimized orthogonal matching pursuit approach.
\newblock {\em IEEE Signal Processing Letters\/}, {\em 9\/}(4), 137--40.

\bibitem[{Rehn \& Sommer(2007)}]{Rehn07}
Rehn, M., \& Sommer, F.~T. (2007).
\newblock A model that uses few active neurones to code visual input predicts
  the diverse shapes of cortical receptive fields.
\newblock {\em Journal of {C}omputational {N}euroscience\/}, {\em 22\/}(2),
  135--46.

\bibitem[{Rissanen(1978)}]{Rissanen78}
Rissanen, J. (1978).
\newblock Modeling by shortest data description.
\newblock {\em Automatica\/}, {\em 14\/}, 465--71.

\bibitem[{Saito(2001)}]{Saito01}
Saito, N. (2001).
\newblock The generalized spike process, sparsity, and statistical
  independence.
\newblock In D.~N. Rockmore, \& D.~M. Healy (Eds.) {\em Modern Signal
  Processing\/}, (p. 317). Cambridge University Press.

\bibitem[{Schemmel et~al.(2006)Schemmel, Gruebl, Meier, \&
  Mueller}]{Schemmel06}
Schemmel, J., Gruebl, A., Meier, K., \& Mueller, E. (2006).
\newblock Implementing synaptic plasticity in a {VLSI} spiking neural network
  model.
\newblock In {IEEE} Press (Ed.) {\em Proceedings of the 2006 International Joint
  Conference on Neural Networks (IJCNN'06)\/}.

\bibitem[{Shahaf et~al.(2008)Shahaf, Eytan, Gal, Kermany, Lyakhov, Zrenner, \&
  Marom}]{Shahaf08}
Shahaf, G., Eytan, D., Gal, A., Kermany, E., Lyakhov, V., Zrenner, C., \&
  Marom, S. (2008).
\newblock Order-based representation in random networks of cortical neurons.
\newblock {\em PLoS Computational Biology\/}, {\em 4\/}(11), e1000228+.

\bibitem[{Smith \& Lewicki(2006)}]{Smith06}
Smith, E.~C., \& Lewicki, M.~S. (2006).
\newblock Efficient auditory coding.
\newblock {\em Nature\/}, {\em 439\/}(7079), 978--82.

\bibitem[{Srinivasan et~al.(1982)Srinivasan, Laughlin, \& Dubs}]{Srinivasan82}
Srinivasan, M.~V., Laughlin, S.~B., \& Dubs, A. (1982).
\newblock Predictive coding: {A} fresh view of inhibition in the retina.
\newblock {\em Proceedings of the Royal Society of London. Series B, Biological
  Sciences\/}, {\em 216\/}(1205), 427--59.

\bibitem[{van Hateren(1993)}]{Hateren93}
van Hateren, J.~H. (1993).
\newblock Spatiotemporal contrast sensitivity of early vision.
\newblock {\em Vision {R}esearch\/}, {\em 33\/}, 257--67.

\bibitem[{Vinje \& Gallant(2000)}]{Vinje00}
Vinje, W.~E., \& Gallant, J.~L. (2000).
\newblock Sparse coding and decorrelation in primary visual cortex during
  natural vision.
\newblock {\em Science\/}, {\em 287\/}, 1273--1276.

\bibitem[{Weber \& Triesch(2008)}]{Weber08}
Weber, C., \& Triesch, J. (2008).
\newblock A sparse generative model of {V}1 simple cells with intrinsic
  plasticity.
\newblock {\em Neural {C}omputation\/}, {\em 20\/}(5), 1261--84.

\bibitem[{Zhaoping(2006)}]{Zhaoping06}
Zhaoping, L. (2006).
\newblock Theoretical understanding of the early visual processes by data
  compression and data selection.
\newblock {\em Network: {C}omputation in {N}eural {S}ystems\/}, {\em 17\/}(4),
  301--34.

\bibitem[{Zibulevsky \& Pearlmutter(2001)}]{Zibulevsky01}
Zibulevsky, M., \& Pearlmutter, B.~A. (2001).
\newblock Blind {S}ource {S}eparation by sparse decomposition.
\newblock {\em Neural {C}omputation\/}, {\em 13\/}(4), 863--82.
\end{thebibliography}
\end{document}